\documentclass[twocolumn,final,10pt]{IEEEtran}
\usepackage{graphicx}

\usepackage{cite}
\usepackage{amsmath}
\usepackage{amsthm}
\usepackage{mathtools}
\usepackage{subfigure}
\usepackage{float}
\usepackage{times}
\usepackage{enumerate}
\usepackage{amssymb}
\usepackage{algorithm}
\usepackage{algorithmic}
\usepackage{multirow}
\usepackage{braket}
\usepackage{multicol}
\usepackage{caption}
\newtheorem{definition}{Definition}

\begin{document}

\title{The Stability Region of the Two-User Interference Channel}

\author{\authorblockN{Nikolaos Pappas\textsuperscript{*}, Marios Kountouris\textsuperscript{*}, Anthony Ephremides\textsuperscript{\ddag}}\\
\authorblockA{\textsuperscript{*} Sup\'{e}lec, Department of Telecommunications, Gif-sur-Yvette, France\\
\textsuperscript{\ddag} Department of Electrical and Computer Engineering and Institute for Systems Research\\
University of Maryland, College Park, MD 20742\\
Email: \{nikolaos.pappas, marios.kountouris\}@supelec.fr, etony@umd.edu
}

\thanks{This work was supported by the Future and Emerging Technologies (FET) project HIATUS within the Seventh Framework Programme for Research of the European Commission under FET-Open grant number: 265578.}}

\maketitle

\begin{abstract}
The stable throughput region of the two-user interference channel is investigated here.
First, the stability region for the general case is characterized. Second, we study the cases where the receivers treat interference as noise or perform successive interference cancelation. Finally, we provide conditions for the convexity/concavity of the stability region and for which a certain interference management strategy leads to broader stability region.
\end{abstract}


\section{Introduction} \label{sec:intro}
Understanding the relationship between information-theoretic capacity and stability regions has received considerable attention in recent years and some progress has been made primarily for multiple access channels. The capacity region is traditionally derived under the assumption of backlogged users and saturated (non-empty) queues. However, under stochastic and bursty traffic arrivals, the maximum stable throughput or stability region (in packets/slot) becomes a meaningful and relevant measure of rates (in packets/slot) in wireless networks. 
The stability region is defined as the union of all arrival rates for which all queue lengths stay finite or have a non-degenerate limiting probability distribution (there are several definitions of stability)~\cite{Szpankowski:stability}.
These two regions are not in general identical and general conditions under which they coincide are known in very few cases.

In this paper, we consider the two-user interference channel, which models communication scenarios in which multiple one-to-one transmissions over a common frequency band are taking place creating interference one each other. The capacity region of the general Gaussian interference channel is a long standing problem and is only known for special cases, such as Gaussian channels with weak (``noisy") or strong interference~\cite{Carleial75, Shang09, AV09}. Furthermore, information-theoretic results advocate for different ways of handling the interference, including orthogonal access, treating interference as noise (IAN), successive interference cancellation (SIC), joint decoding and interference alignment~\cite{b:Gamal_NIT}. Here, we investigate the stability region of the two-user interference channel, which, to the best of our knowledge, has not been reported to the literature.
In~\cite{b:Naware}, the  effect of multipacket reception on stability and delay of slotted ALOHA-based random access systems is considered. In ~\cite{b:Simeone}, the authors studied a cognitive interference channel, as well as the case of a primary user and a cognitive user with and without relaying capabilities. The maximum stable throughput of the cognitive user for a fixed throughput selected by the primary user is derived.

In this work, we investigate the two-user interference channel, where each user has bursty arrivals and transmits a packet whenever its queue is not empty, and we obtain the exact stability region for the general case. The characterization of the stability region is a challenging problem due to the fact that the user queues are coupled, i.e. the service process of a queue depends on the status of the other queues. To overcome this difficulty, the stochastic dominance technique is used here~\cite{rao:stability}.
We also consider the cases where each receiver treats interference as noise or employ successive interference cancelation. Finally, we derive conditions for the shape of the stability region (concave or convex) and we show under which system parameters, each interference management technique is superior (in the sense of broader stability region) compared to the other.

\section{System Model} \label{sec:model}

\begin{figure}[t]
\centering
\includegraphics[scale=1.1]{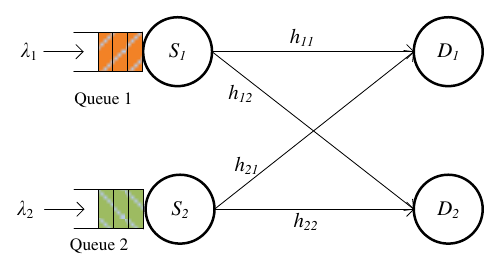}
\caption{Two-user interference channel with bursty arrivals.}
\label{fig:model}
\end{figure}

We consider a two-user interference channel, as depicted in Fig.~\ref{fig:model}, in which each source $S_i,i=1,2$ intends to communicate with its respective destination $D_{i},i=1,2$. The packet arrival processes at $S_1$ and $S_2$ are assumed to be independent and stationary with mean rates $\lambda_1$ and $\lambda_2$, respectively. Transmitter $S_i$ has an infinite capacity queue to store incoming packets and $Q_i$ denotes the size in number packets of the $i$-th queue. The transmission rates of $S_1$ and $S_2$ are fixed at $R_1$ and $R_2$, respectively.

Time is assumed to be slotted and each source transmits a packet in a timeslot if its queue is not empty; otherwise it remains silent.
The transmission of one packet requires one timeslot and we assume that ACKs are instantaneous and error-free.
A block fading channel model is considered here with Rayleigh fading, i.e. the fading coefficients $h_{ij}$ remain constant during one timeslot, but change independently from one timeslot to another based on a circularly symmetric complex Gaussian distribution with zero mean and unit variance. The noise is assumed to be additive white Gaussian with zero mean and unit variance. With $p_i$ we denote the transmission power of source $S_i$, and $r_{ij}$ is the distance between transmitter $S_i$ and receiver $D_j$ with $a$ being the path loss exponent.

Let $\mathcal{D}^\mathcal{T}_{i}$ denote the event that destination $i$ is able to decode the packet transmitted from the $i$-th source given a set of active transmitters denoted by $\mathcal{T}$ i.e. $\mathcal{D}^{\{ 1,2 \} }_{1}$ denotes the event that the first destination can decode the information from the first source when both transmitters are active ($\mathcal{T} = \{1,2\}$). When only $S_i$ is active the event $\mathcal{D}^{\{ i \} }_{i}$ is defined as
\begin{equation}
\mathcal{D}^{\{ i \} }_{i} \triangleq \left\lbrace R_i \leq \log_2 \left(1+ |h_{ii}|^2 r^{-a}_{ii} p_i \right) \right\rbrace,
\end{equation}
which is equivalent to $\mathcal{D}^{\{ i \} }_{i} = \left\lbrace 2^{R_i} - 1 \leq |h_{ii}|^2 r^{-a}_{ii} p_i \right\rbrace$.


For convenience we define $ \mathrm{SNR}_i \triangleq |h_{ii}|^2 r^{-a}_{ii} p_i$ and $\gamma_i \triangleq 2^{R_i} - 1$. The probability that the link $ii$ is not in outage when only $S_i$ is active is given by~\cite{b:Tse}
\begin{equation} \label{eq:SNR}
\mathrm{Pr}\left(\mathcal{D}^{\{i\}}_{i}\right) =\mathrm{Pr} \left\lbrace \mathrm{SNR}_i \geq \gamma_i \right\rbrace = \exp \left(- \frac{\gamma_i r^{a}_{ii}}{p_{i}}\right).
\end{equation}
In the next sections, the events $\mathcal{D}^{\{ i,j \} }_{i}$ (both sources are active) are defined based on the specific interference treatment on each receiver.

%
%
%

We adopt the definition of queue stability used in~\cite{Szpankowski:stability}.

\begin{definition}
Denote by $Q_i^t$ the length of queue $i$ at the beginning of time slot $t$. The queue is said to be \emph{stable} if
%
\begin{equation}\label{eqn:definition_stability}
    \lim_{t \rightarrow \infty} {Pr}[Q_i^t < {x}] = F(x)  \text{ and } \lim_{ {x} \rightarrow \infty} F(x) = 1.
\end{equation}

If $\lim_{x \rightarrow \infty}  \lim_{t \rightarrow \infty} \inf {Pr}[Q_i^t < {x}] = 1$, the queue is \emph{substable}. If a queue is stable, then it is also substable. If a queue is not substable, then we say it is unstable.

%
\end{definition}

Loynes' theorem~\cite{b:Loynes} states that if the arrival and service processes of a queue are strictly jointly stationary and the average arrival rate is less than the average service rate, then the queue is stable. The stability region of the system is defined as the set of arrival rate vectors $(\lambda_1, \lambda_2)$ for which the queues in the system are stable.


\section{Stability Analysis: General Case} \label{sec:general_region}

In this section we provide the stability region in a parametric form without considering any specific technique for treating the interference at the receivers. The service rates for the sources are given by
\begin{equation} \label{eq:mu_1}
\mu_1 = \mathrm{Pr} [Q_2 > 0] \mathrm{Pr}\left(\mathcal{D}^{\{1,2\}}_{1}\right) + \mathrm{Pr} [Q_2 = 0] \mathrm{Pr}\left(\mathcal{D}^{\{1\}}_{1}\right)
\end{equation}

\vspace{-5mm}

\begin{equation} \label{eq:mu_2}
\mu_2 = \mathrm{Pr} [Q_1 > 0] \mathrm{Pr}\left(\mathcal{D}^{\{1,2\}}_{2}\right) + \mathrm{Pr} [Q_1 = 0] \mathrm{Pr}\left(\mathcal{D}^{\{2\}}_{2}\right).
\end{equation}


Since the average service rate of each queue depends on the queue size of the other queues, it cannot be computed directly. Therefore, we apply the stochastic dominance technique~\cite{rao:stability}; that is, we construct hypothetical dominant systems, in which one of the sources transmits dummy packets when its packet queue is empty, while the other transmits according to its traffic.


\subsection{The first dominant system}

We consider the first dominant system, in which $S_1$ transmits dummy packets whenever its queue is empty, while $S_2$ behaves in the same way as in the original system. All other assumptions remain unaltered in the dominant system.

From Loyne's criterion~\cite{b:Loynes} it is known that the queue at the second source is stable if and only if $\lambda_2 < \mu_2$, where $\mu_i$ is the service rate for source $S_i, i=1,2$. Therefore, the stability condition is given by

\begin{equation}  \label{eq:stable2_dom1}
\lambda_2 < \mu_2 = \mathrm{Pr}\left(\mathcal{D}^{\{1,2\}}_{2}\right).
\end{equation}

From Little's theorem~\cite{b:Bertsekas}, the probability that the queue of the second transmitter is empty is given by

\begin{equation} \label{eq:q2nonempty_dom1}
\mathrm{Pr}[Q_2 > 0]=\frac{\lambda_2}{\mathrm{Pr}\left(\mathcal{D}^{\{1,2\}}_{2}\right)}.
\end{equation}

Substituting (\ref{eq:q2nonempty_dom1}) into (\ref{eq:mu_1}), we have that the service rate for the first source is given by

\begin{equation} \label{eq:mu1_dom1}
\mu_1 = \mathrm{Pr}\left(\mathcal{D}^{\{1\}}_{1} \right) - \frac{\lambda_2 \mathrm{Pr}\left(\mathcal{D}^{\{1\}}_{1}\right)}{\mathrm{Pr}\left(\mathcal{D}^{\{1,2\}}_{2}\right)} + \frac{\lambda_2\mathrm{Pr}\left(\mathcal{D}^{\{1,2\}}_{1}\right)}{\mathrm{Pr}\left(\mathcal{D}^{\{1,2\}}_{2}\right)}.
\end{equation}

The queue at the first source is stable if and only if $\lambda_1 < \mu_1$; hence the stability condition is given by
\begin{equation} \label{eq:stable1_dom1}
\lambda_1 < \mathrm{Pr}\left(\mathcal{D}^{\{1\}}_{1}\right) - \frac{\lambda_2 \mathrm{Pr}\left(\mathcal{D}^{\{1\}}_{1}\right)}{\mathrm{Pr}\left(\mathcal{D}^{\{1,2\}}_{2}\right)} + \frac{\lambda_2\mathrm{Pr}\left(\mathcal{D}^{\{1,2\}}_{1}\right)}{\mathrm{Pr}\left(\mathcal{D}^{\{1,2\}}_{2}\right)}.
\end{equation}



The stability region $\mathcal{R}_1$, obtained by the first dominant system and conditions (\ref{eq:stable1_dom1}) and (\ref{eq:stable2_dom1}), is given by (\ref{eq:R_1}) (on top of the next page).

\begin{figure*}[!t]
\begin{equation} \label{eq:R_1}
\mathcal{R}_1 = \left\lbrace (\lambda_{1},\lambda_{2}): \frac{\lambda_1}{\mathrm{Pr}\left(\mathcal{D}^{\{1\}}_{1}\right)} + \frac{\left[\mathrm{Pr}\left(\mathcal{D}^{\{1\}}_{1}\right) - \mathrm{Pr}\left(\mathcal{D}^{\{1,2\}}_{1}\right) \right] \lambda_2}{\mathrm{Pr}\left(\mathcal{D}^{\{1\}}_{1}\right)\mathrm{Pr}\left(\mathcal{D}^{\{1,2\}}_{2}\right)}   <  1,
\lambda_2 < \mathrm{Pr}\left(\mathcal{D}^{\{1,2\}}_{2}\right)  \right\rbrace
\end{equation}
\begin{equation} \label{eq:R_2}
\mathcal{R}_2 = \left\lbrace (\lambda_{1},\lambda_{2}): \frac{\lambda_2}{\mathrm{Pr}\left(\mathcal{D}^{\{2\}}_{2}\right)} + \frac{\left[\mathrm{Pr}\left(\mathcal{D}^{\{2\}}_{2}\right) - \mathrm{Pr}\left(\mathcal{D}^{\{1,2\}}_{2}\right) \right] \lambda_1}{\mathrm{Pr}\left(\mathcal{D}^{\{2\}}_{2}\right)\mathrm{Pr}\left(\mathcal{D}^{\{1,2\}}_{1}\right)}   <  1,
\lambda_1 < \mathrm{Pr}\left(\mathcal{D}^{\{1,2\}}_{1}\right)  \right\rbrace
\end{equation}
\end{figure*}



An important observation made in \cite{rao:stability} is that the stability conditions obtained by the stochastic dominance technique are not only sufficient but also necessary conditions for the stability of the original system. The \emph{indistinguishability} argument~\cite{rao:stability} applies to our problem as well. Based on the construction of the dominant system, it is easy to see that the queues of the dominant system are always larger in size than those of the original system, provided they are both initialized to the same value. Therefore, given $\lambda_{2}<\mu_{2}$, if for some $\lambda_{1}$, the queue at $S_1$ is stable in the dominant system, then the corresponding queue in the original system must be stable. Conversely, if for some $\lambda_{1}$ in the dominant system, the queue at node $S_1$ saturates, then it will not transmit dummy packets, and as long as $S_1$ has a packet to transmit, the behavior of the dominant system is identical to that of the original system because dummy packet transmissions are eliminated as we approach the stability boundary. Therefore, the original and the dominant system are indistinguishable at the boundary points.

\subsection{The second dominant system}

In the second dominant system, source $S_2$ transmits dummy packets when its queue is empty and all other assumptions remain unaltered.
Similarly to the first dominant system and using Loyne's criterion, the stability condition is given by
\begin{equation}  \label{eq:stable1_dom2}
\lambda_1 < \mu_1 = \mathrm{Pr}\left(\mathcal{D}^{\{1,2\}}_{1}\right),
\end{equation}
since the queue at the first source is stable if and only if $\lambda_1 < \mu_1$.

The probability that the queue of the first user is empty is given by
\begin{equation} \label{eq:q1nonempty_dom2}
\mathrm{Pr}[Q_1 > 0]=\frac{\lambda_1}{\mathrm{Pr}\left(\mathcal{D}^{\{1,2\}}_{1}\right)}.
\end{equation}

Therefore, substituting (\ref{eq:q1nonempty_dom2}) into (\ref{eq:mu_2}) and given that the second queue is stable if and only if $\lambda_2 < \mu_2$, the stability condition is given by
\begin{equation} \label{eq:stable2_dom2}
\lambda_2 < \mu_2 = \mathrm{Pr}\left(\mathcal{D}^{\{2\}}_{2} \right) - \frac{\lambda_1 \mathrm{Pr}\left(\mathcal{D}^{\{2\}}_{2}\right)}{\mathrm{Pr}\left(\mathcal{D}^{\{1,2\}}_{1}\right)} + \frac{\lambda_1\mathrm{Pr}\left(\mathcal{D}^{\{1,2\}}_{2}\right)}{\mathrm{Pr}\left(\mathcal{D}^{\{1,2\}}_{1}\right)}.
\end{equation}

%
The stability region $\mathcal{R}_2$, obtained by the second dominant system and conditions (\ref{eq:stable2_dom2}) and (\ref{eq:stable1_dom2}), is given on the top of this page by (\ref{eq:R_2}).


Similarly to the first dominant system, the indistinguishability argument also holds as approaching the boundary points (saturation).

%

\begin{figure}[t]
\centering
\includegraphics[scale=0.58]{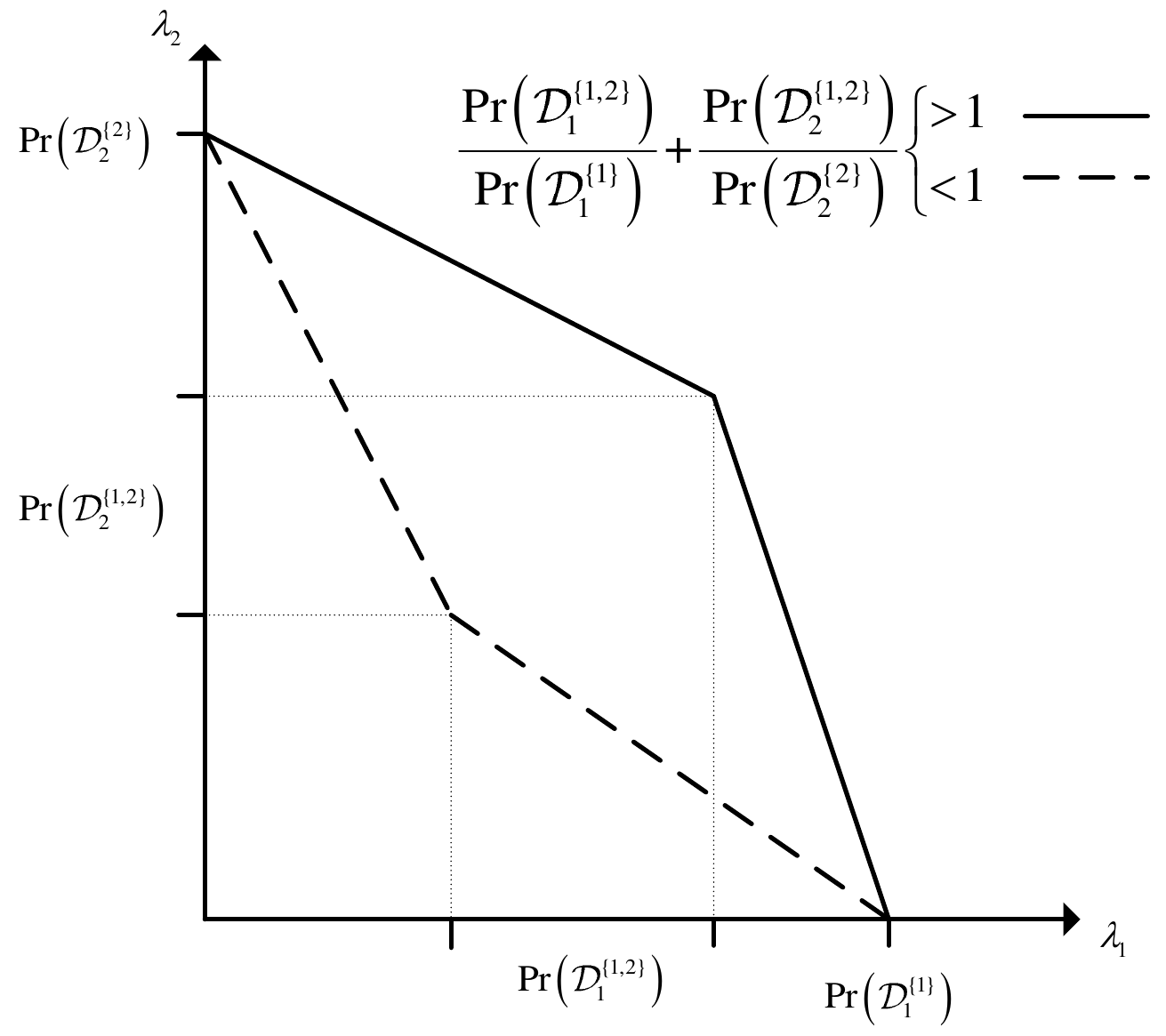}
\caption{The stability region for the general case.}
\label{fig:region}
\end{figure}

The stability region is given by $\mathcal{R} = \mathcal{R}_1 \bigcup \mathcal{R}_2$ and is depicted in Fig.~\ref{fig:region}.
It is easy to see that if
\begin{equation}\label{eq:convexity_general}
\frac{\mathrm{Pr}\left(\mathcal{D}^{\{1,2\}}_{1}\right)}{\mathrm{Pr}\left(\mathcal{D}^{\{1\}}_{1}\right)}+
\frac{\mathrm{Pr}\left(\mathcal{D}^{\{1,2\}}_{2}\right)}{\mathrm{Pr}\left(\mathcal{D}^{\{2\}}_{2}\right)} \gtrless 1,
\end{equation}
then the stability region is concave/convex.

The aforementioned result on the stability region holds for any interference management technique. In the following sections, we particularize the stability conditions considering different ways of treating the interference.

\section{IAN at Both Receivers} \label{sec:IAN_region}

We provide here the stability region when both destinations decode their individual messages by treating the interference from unintended sources as noise. When both transmitters are active, the event $\mathcal{D}^{\{ i,j \} }_{i}$ is given by

\begin{equation}
\mathcal{D}^{\{ i,j \} }_{i} \triangleq \left\lbrace R_i \leq \log_2 \left(1+ \frac{|h_{ii}|^2 r^{-a}_{ii} p_i}{1+|h_{ji}|^2 r^{-a}_{ji} p_j} \right) \right\rbrace ,
\end{equation}
which is equivalent to
\begin{equation}
\mathcal{D}^{\{ i,j \} }_{i} = \left\lbrace \gamma_i \leq \frac{|h_{ii}|^2 r^{-a}_{ii} p_i}{1+|h_{ji}|^2 r^{-a}_{ji} p_j} \triangleq \mathrm{SINR}_i \right\rbrace.
\end{equation}

The probability that the channel $ii$ is not in outage when all the sources are active is given by~\cite{b:Tse}:
\begin{equation} \label{eq:SINR_IAN}
\begin{aligned}
\mathrm{Pr}\left(\mathcal{D}^{\{i,j\}}_{i}\right) =\mathrm{Pr} \left\lbrace \mathrm{SINR}_i \geq \gamma_i \right\rbrace = \\ = \exp \left(- \frac{\gamma_i r^{a}_{ii}}{p_{i}} \right) \left[1+\gamma_i \frac{p_{j}}{p_{i}} \left( \frac{r_{ii}}{r_{ji}} \right)^a \right]^{-1} ,\text{ for }i=1,2
\end{aligned}
\end{equation}

Substituting (\ref{eq:SNR}), (\ref{eq:SINR_IAN}) to (\ref{eq:R_1}), (\ref{eq:R_2}), we obtain the stability sub-regions (\ref{eq:R_1_IAN}) and (\ref{eq:R_2_IAN}) respectively, given on the top of next page.
\begin{figure*}
\begin{equation} \label{eq:R_1_IAN}
\mathcal{R}^{\mathrm{IAN}}_1= \left\lbrace  (\lambda_{1},\lambda_{2}): \frac{\lambda_1}{\exp \left(- \frac{\gamma_1 r^{a}_{11}}{p_{1}}\right)} + \frac{\gamma_1 \frac{p_2}{p_1} \left( \frac{r_{11}}{r_{21}} \right)^a + \gamma_1 \gamma_2 \left( \frac{r_{11} r_{22}}{r_{12} r_{21}} \right)^a}{\exp \left(- \frac{\gamma_2 r^{a}_{22}}{p_{2}} \right)} \lambda_2 < 1, \lambda_2 < \frac{\exp \left(- \frac{\gamma_2 r^{a}_{22}}{p_{2}} \right)}{ \left[1+\gamma_2 \frac{p_{1}}{p_{2}} \left( \frac{r_{22}}{r_{12}} \right)^a \right]}  \right\rbrace
\end{equation}
\begin{equation} \label{eq:R_2_IAN}
\mathcal{R}^{\mathrm{IAN}}_2= \left\lbrace  (\lambda_{1},\lambda_{2}): \frac{\lambda_2}{\exp \left(- \frac{\gamma_2 r^{a}_{22}}{p_{2}}\right)} + \frac{\gamma_2 \frac{p_1}{p_2} \left( \frac{r_{22}}{r_{12}} \right)^a + \gamma_1 \gamma_2 \left( \frac{r_{22} r_{11}}{r_{12} r_{21}} \right)^a}{\exp \left(- \frac{\gamma_1 r^{a}_{11}}{p_{1}} \right)} \lambda_1 < 1, \lambda_1 < \frac{\exp \left(- \frac{\gamma_1 r^{a}_{11}}{p_{1}} \right)}{ \left[1+\gamma_1 \frac{p_{2}}{p_{1}} \left( \frac{r_{11}}{r_{21}} \right)^a \right]}  \right\rbrace
\end{equation}
\begin{equation}\label{eq:R_1_IAN_C}
\mathcal{R}^{\mathrm{IAN}}_1= \left\lbrace  (\lambda_{1},\lambda_{2}): \frac{\lambda_1}{\mathrm{Pr} \left\lbrace \mathrm{SNR}_1 \geq \gamma_1 \right\rbrace
} + \frac{\mathrm{Pr} \left\lbrace \mathrm{SNR}_1 \geq \gamma_1 \right\rbrace - \mathrm{Pr} \left\lbrace \mathrm{SINR}_1 \geq \gamma_1 \right\rbrace}{\mathrm{Pr} \left\lbrace \mathrm{SNR}_1 \geq \gamma_1 \right\rbrace \mathrm{Pr} \left\lbrace \mathrm{SINR}_2 \geq \gamma_2 \right\rbrace} \lambda_2 < 1, \lambda_2 < \mathrm{Pr} \left\lbrace \mathrm{SINR}_2 \geq \gamma_2 \right\rbrace  \right\rbrace
\end{equation}
\begin{equation}\label{eq:R_2_IAN_C}
\mathcal{R}^{\mathrm{IAN}}_2= \left\lbrace  (\lambda_{1},\lambda_{2}): \frac{\lambda_2}{\mathrm{Pr} \left\lbrace \mathrm{SNR}_2 \geq \gamma_2 \right\rbrace
} + \frac{\mathrm{Pr} \left\lbrace \mathrm{SNR}_2 \geq \gamma_2 \right\rbrace - \mathrm{Pr} \left\lbrace \mathrm{SINR}_2 \geq \gamma_2 \right\rbrace}{\mathrm{Pr} \left\lbrace \mathrm{SNR}_2 \geq \gamma_2 \right\rbrace \mathrm{Pr} \left\lbrace \mathrm{SINR}_1 \geq \gamma_1 \right\rbrace} \lambda_1 < 1, \lambda_1 < \mathrm{Pr} \left\lbrace \mathrm{SINR}_1 \geq \gamma_1 \right\rbrace  \right\rbrace
\end{equation}
\end{figure*}
The $\mathcal{R}^{\mathrm{IAN}}_1$ and $\mathcal{R}^{\mathrm{IAN}}_2$ can be presented in a more compact form given in (\ref{eq:R_1_IAN_C}) and (\ref{eq:R_2_IAN_C}), respectively.


The stability region when both receivers can decode their messages by treating interference as noise is $\mathcal{R}^{\mathrm{IAN}} = \mathcal{R}^{\mathrm{IAN}}_1 \cup \mathcal{R}^{\mathrm{IAN}}_2$.

Substituting (\ref{eq:SNR}) and (\ref{eq:SINR_IAN}) to (\ref{eq:convexity_general}) we have that $\mathcal{R}^{\mathrm{IAN}}$ is convex/concave when $\frac{\mathrm{Pr}\left\lbrace\mathrm{SINR}_1 \geq \gamma_1 \right\rbrace}{\mathrm{Pr}\left\lbrace\mathrm{SNR}_1 \geq \gamma_1\right\rbrace} + \frac{\mathrm{Pr}\left\lbrace\mathrm{SINR}_2 \geq \gamma_2 \right\rbrace}{\mathrm{Pr}\left\lbrace\mathrm{SNR}_2 \geq \gamma_2 \right\rbrace} \gtrless 1$, which leads to the condition

\begin{equation} \label{eq:convexity_IAN}
\gamma_1 \gamma_2 \lessgtr \left( \frac{r_{12} r_{21}}{r_{22} r_{11}} \right)^a.
\end{equation}

\begin{figure}[t]
\centering
\includegraphics[scale=0.65]{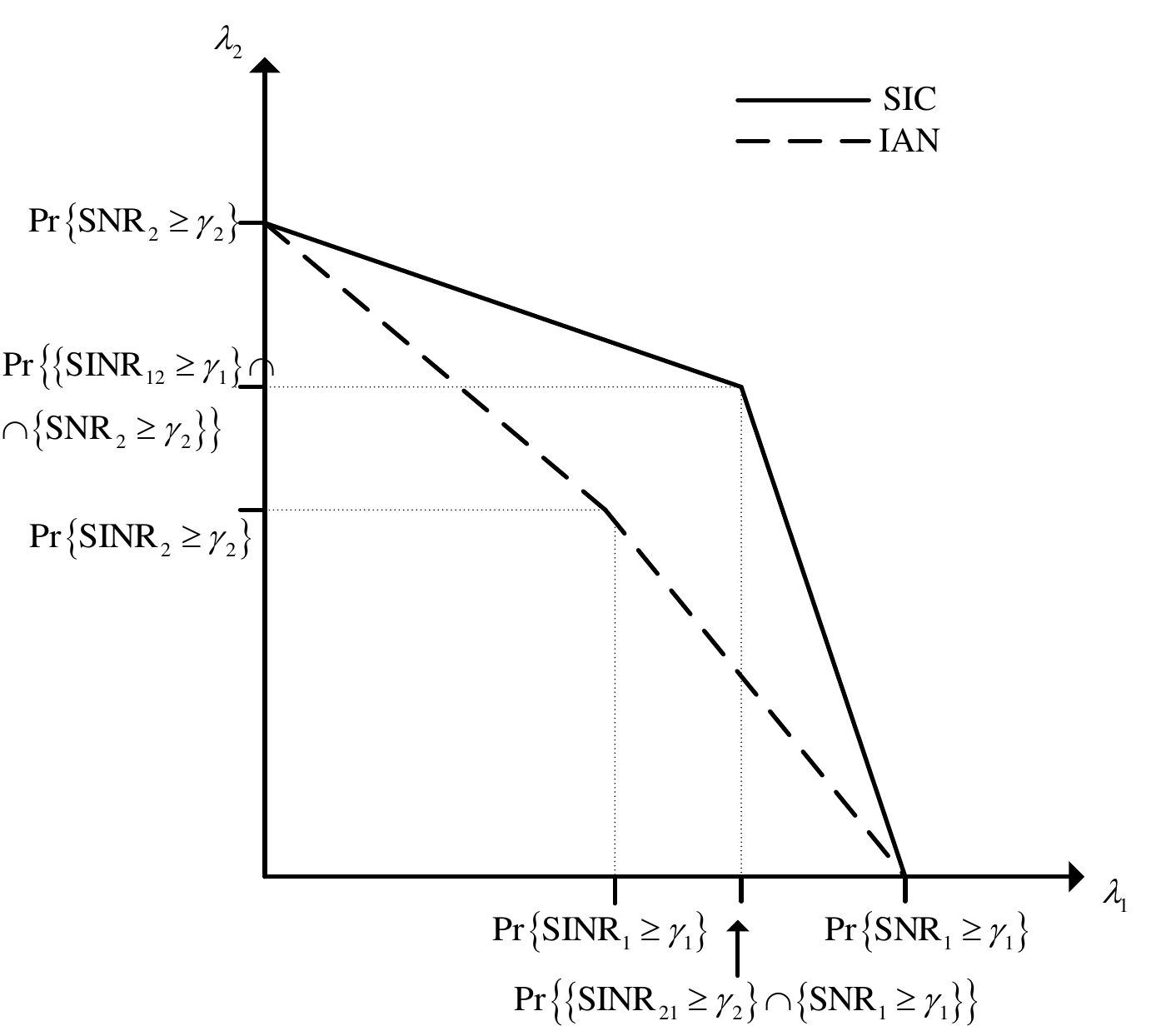}
\caption{The concave stability region for the case where both receivers apply IAN and SIC. The condition (\ref{eq:SIC_optimal_condition0}) holds for both receivers.}
\label{fig:SIC_vs_IAN}
\end{figure}

The $\mathcal{R}^{\mathrm{IAN}}$ for the concave case is depicted in Fig.~\ref{fig:SIC_vs_IAN}.

\section{SIC at both receivers} \label{sec:SIC_region}

In this section we derive the stability region when both receivers employ successive interference cancelation when both transmitters are active. If the destination $D_j$ knows the codebook of $S_i$, it can perform SIC by first decoding the message sent by $S_i$, removing its contribution (interference) to the received signal, and then decoding its own message.

The receiver $D_i$ is able to decode the interference (when both sources are active) if the following conditions are satisfied
\begin{align}
R_j \leq \log_2 \left(1+ \frac{|h_{ji}|^2 r^{-a}_{ji} p_j}{1+|h_{ii}|^2 r^{-a}_{ii} p_i} \right), \\
R_i \leq \log_2 \left(1+ |h_{ii}|^2 r^{-a}_{ii} p_i \right),
\end{align}
which are equivalent to
\begin{align}
\gamma_j=2^{R_j} -1 \leq \frac{|h_{ji}|^2 r^{-a}_{ji} p_j}{1+|h_{ii}|^2 r^{-a}_{ii} p_i} \triangleq \mathrm{SINR}_{ji}\text{ and } \gamma_i \leq \mathrm{SNR}_i.
\end{align}

The event $\mathcal{D}^{\{i,j\}}_{i}$ is given by $\mathcal{D}^{\{i,j\}}_{i} = \left\lbrace \mathrm{SINR}_{ji} \geq \gamma_j \right\rbrace \cap \left\lbrace \mathrm{SNR}_i \geq \gamma_i \right\rbrace$, and the probability that $D_i$ can decode the transmitted information from $S_i$ (given that both sources all active) is

\begin{equation}\label{eq:SINR_SIC}
\begin{aligned}
\mathrm{Pr}\left(\mathcal{D}^{\{i,j\}}_{i}\right) =\mathrm{Pr}\left\lbrace \left\lbrace \mathrm{SINR}_{ji} \geq \gamma_j \right\rbrace \cap \left\lbrace \mathrm{SNR}_i \geq \gamma_i \ \right\rbrace \right\rbrace \\
= \exp \left(- \frac{\gamma_i r^{a}_{ii}}{p_{i}} \right)  \exp \left[- \frac{\gamma_j(1+\gamma_i) r^{a}_{ji}}{p_{j}} \right]  \left[1+\gamma_j \frac{p_{i}}{p_{j}} \left( \frac{r_{ji}}{r_{ii}} \right)^a \right]^{-1}.
\end{aligned}
\end{equation}

The probability $\mathrm{Pr}(\mathcal{D}^{\{i\}}_{i}) = \mathrm{Pr}(\mathrm{SNR}_i \geq \gamma_i)$ is given by (\ref{eq:SNR}).
Substituting (\ref{eq:SNR}) and (\ref{eq:SINR_SIC}) to (\ref{eq:R_1}) and (\ref{eq:R_2}) we obtain that the subregions are (\ref{eq:R_1_SIC_C}) and (\ref{eq:R_2_SIC_C}).

\begin{figure*}
{\footnotesize
\begin{equation}\label{eq:R_1_SIC_C}
\mathcal{R}^{\mathrm{SIC}}_1= \left\lbrace  (\lambda_{1},\lambda_{2}): \frac{\lambda_1}{\mathrm{Pr} \left\lbrace \mathrm{SNR}_1 \geq \gamma_1 \right\rbrace
} + \frac{1-\mathrm{Pr}\left\lbrace \left\lbrace \mathrm{SINR}_{21} \geq \gamma_2 \right\rbrace \mid \left\lbrace \mathrm{SNR}_1 \geq \gamma_1 \right\rbrace \right\rbrace}{\mathrm{Pr}\left\lbrace \left\lbrace \mathrm{SINR}_{12} \geq \gamma_1 \right\rbrace \cap \left\lbrace \mathrm{SNR}_2 \geq \gamma_2 \right\rbrace \right\rbrace} \lambda_2 < 1, \lambda_2 < \mathrm{Pr}\left\lbrace \left\lbrace \mathrm{SINR}_{12} \geq \gamma_1 \right\rbrace \cap \left\lbrace \mathrm{SNR}_2 \geq \gamma_2 \right\rbrace \right\rbrace  \right\rbrace.
\end{equation}
\begin{equation}\label{eq:R_2_SIC_C}
\mathcal{R}^{\mathrm{SIC}}_2= \left\lbrace  (\lambda_{1},\lambda_{2}): \frac{\lambda_2}{\mathrm{Pr} \left\lbrace \mathrm{SNR}_2 \geq \gamma_2 \right\rbrace
} + \frac{1-\mathrm{Pr}\left\lbrace \left\lbrace \mathrm{SINR}_{12} \geq \gamma_1 \right\rbrace \mid \left\lbrace \mathrm{SNR}_2 \geq \gamma_2 \right\rbrace \right\rbrace}{\mathrm{Pr}\left\lbrace \left\lbrace \mathrm{SINR}_{21} \geq \gamma_2 \right\rbrace \cap \left\lbrace \mathrm{SNR}_1 \geq \gamma_1 \right\rbrace \right\rbrace} \lambda_1 < 1, \lambda_1 < \mathrm{Pr}\left\lbrace \left\lbrace \mathrm{SINR}_{21} \geq \gamma_2 \right\rbrace \cap \left\lbrace \mathrm{SNR}_1 \geq \gamma_1 \right\rbrace \right\rbrace  \right\rbrace.
\end{equation}
}
\end{figure*}




The stability region $\mathcal{R}^{\mathrm{SIC}} = \mathcal{R}^{\mathrm{SIC}}_1 \cup \mathcal{R}^{\mathrm{SIC}}_2 $ is concave/convex if 
\begin{align*}
\mathrm{Pr}\left\lbrace \left\lbrace \mathrm{SINR}_{21} \geq \gamma_2 \right\rbrace \mid \left\lbrace \mathrm{SNR}_1 \geq \gamma_1 \ \right\rbrace \right\rbrace + \\ \mathrm{Pr}\left\lbrace \left\lbrace \mathrm{SINR}_{12} \geq \gamma_1 \right\rbrace \mid \left\lbrace \mathrm{SNR}_2 \geq \gamma_2 \ \right\rbrace \right\rbrace \gtrless 1.
\end{align*}

The $\mathcal{R}^{\mathrm{SIC}}$ for the concave case is depicted in Fig.~\ref{fig:SIC_vs_IAN}.
%


\subsection{SIC vs. IAN}
We provide here the conditions under which SIC is better than IAN in the sense that $R^{\mathrm{IAN}} \subset R^{\mathrm{SIC}}$.
Comparing $R^{\mathrm{IAN}}$ and $R^{\mathrm{SIC}}$, we have that SIC provides better performance when the following condition is met for both receivers (see Fig.~\ref{fig:SIC_vs_IAN}):

\begin{equation}\label{eq:SIC_optimal_condition0}
\mathrm{Pr} \left\lbrace \mathrm{SINR}_{i} \geq \gamma_i \right\rbrace <\mathrm{Pr}\left\lbrace \left\lbrace \mathrm{SINR}_{ji} \geq \gamma_j \right\rbrace \cap \left\lbrace \mathrm{SNR}_i \geq \gamma_i \ \right\rbrace \right\rbrace ,
\end{equation}
which leads to the following condition after substitution:
\begin{equation} \label{eq:SIC_optimal_condition}
\frac{1+\gamma_j \frac{p_{i}}{p_{j}} \left( \frac{r_{ji}}{r_{ii}} \right)^a}{1+\gamma_i \frac{p_{j}}{p_{i}} \left( \frac{r_{ii}}{r_{ji}} \right)^a} < \exp \left(- \frac{\gamma_j (1+\gamma_i) r^{a}_{ji}}{p_{j}} \right).
\end{equation}

If the condition (\ref{eq:SIC_optimal_condition}) is not satisfied at both receivers, then IAN provides superior performance as compared to SIC. In the case that the condition is not met at $D_i$ but holds for $D_j$, then IAN should be used for $D_i$ and SIC for $D_j$. The stability region for the aforementioned case is provided in the next section.


\section{SIC at the First RX - IAN at the Second RX} \label{sec:SICIAN_region}
In this section we consider the case where, without loss of generality, the first receiver decodes the interference using SIC (condition (\ref{eq:SIC_optimal_condition}) holds for $D_1$), whereas the second receiver applies IAN, i.e. inequality (\ref{eq:SIC_optimal_condition}) holds with the opposite direction.


For the first destination, which decodes the transmitted message applying SIC, the probabilty of successful event $\mathrm{Pr}\left(\mathcal{D}^{\{1,2\}}_{1}\right) = \mathrm{Pr}\left\lbrace \left\lbrace\mathrm{SNR}_1 \geq \gamma_1\right\rbrace \cap \mathrm{Pr}\left\lbrace\mathrm{SINR}_{21} \geq \gamma_2\right\rbrace \right\rbrace,$ is given by (\ref{eq:SINR_SIC}).
For the second destination, the probability $\mathrm{Pr}\left(\mathcal{D}^{\{1,2\}}_{2}\right) =\mathrm{Pr} \left\lbrace  \mathrm{SINR}_2 \geq \gamma_2 \right\rbrace $ is given by (\ref{eq:SINR_IAN}).
Note that when only $i$-th source transmits, then we need that the SNR to be greater than threshold $\gamma_i$.

\begin{figure}[t]
\centering
\includegraphics[scale=0.65]{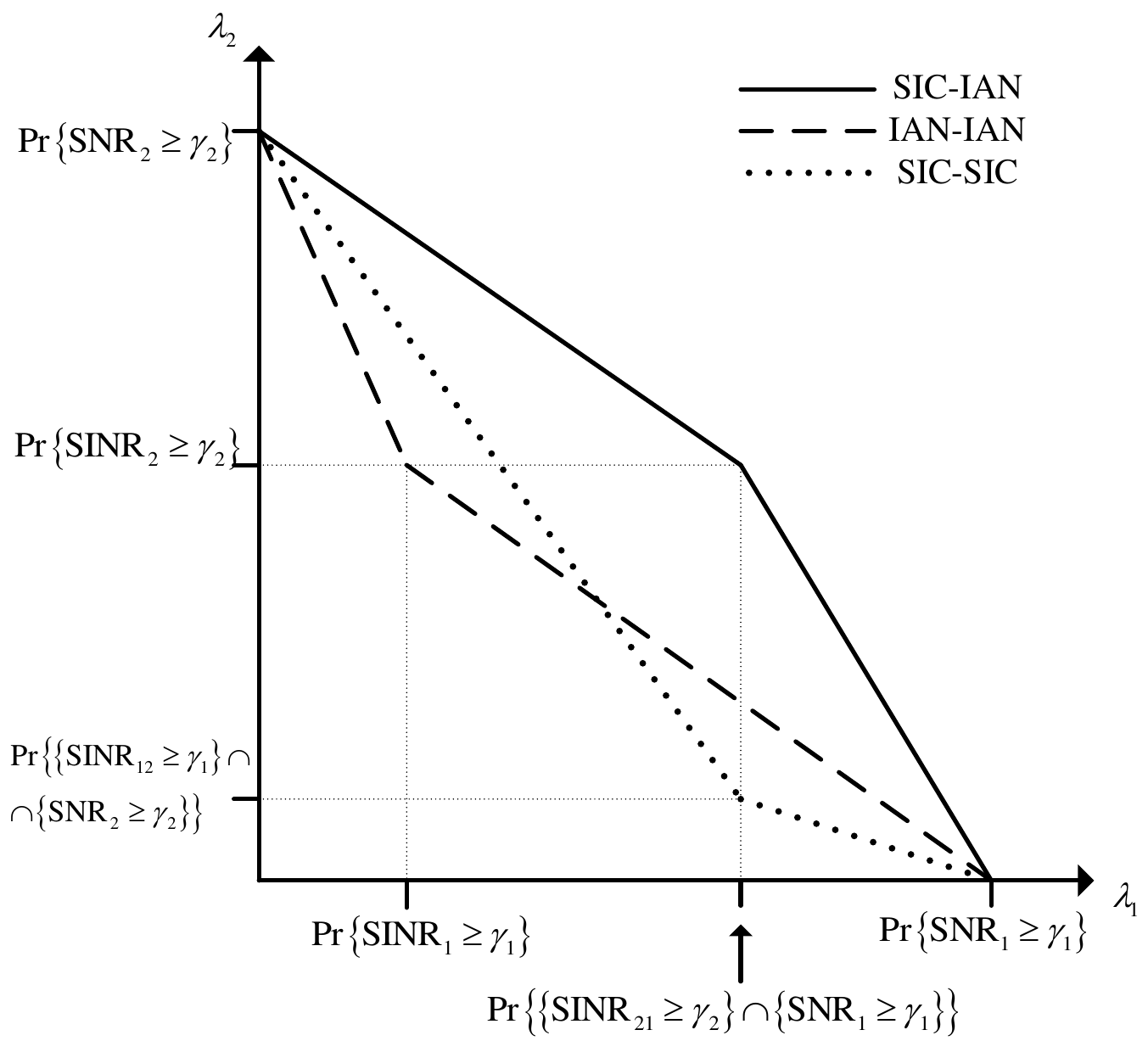}
\caption{Stability region comparison when condition (\ref{eq:SIC_optimal_condition0}) does not hold for both receivers.}
\label{fig:SIC_IAN}
\end{figure}

The stability region is given by $\mathcal{R}^{\mathrm{SIC-IAN}}= \mathcal{R}^{\mathrm{SIC-IAN}}_1 \cup \mathcal{R}^{\mathrm{SIC-IAN}}_2 $ and is shown in Fig.~\ref{fig:SIC_IAN}. The subregions (\ref{eq:R_1_IANSIC}) and (\ref{eq:R_2_IANSIC}) are obtained respectively by substituting (\ref{eq:SINR_IAN}) and (\ref{eq:SINR_SIC}) into (\ref{eq:R_1}) and (\ref{eq:R_2}) for the first and section destination respectively.

\begin{equation} \label{eq:R_1_IANSIC}
\begin{aligned}
\mathcal{R}^{\mathrm{SIC-IAN}}_1= \left\{  (\lambda_{1},\lambda_{2}):
\frac{\lambda_1}{\mathrm{Pr} \left\lbrace\mathrm{SNR}_{1} \geq \gamma_1 \right\rbrace}+ \right. \\
\left. + \frac{1-\mathrm{Pr}\left\lbrace \left\lbrace \mathrm{SINR}_{21} \geq \gamma_2 \right\rbrace \mid \left\lbrace \mathrm{SNR}_1 \geq \gamma_1 \right\rbrace \right\rbrace}{\mathrm{Pr}\left\lbrace\mathrm{SINR}_{2} \geq \gamma_2 \right\rbrace}\lambda_2 < 1, \right. \\
\left. \lambda_2 < \mathrm{Pr}\left\lbrace\mathrm{SINR}_{2} \geq \gamma_2 \right\rbrace \right\}
\end{aligned}
\end{equation}
{\scriptsize
\begin{equation}
\begin{aligned} \label{eq:R_2_IANSIC}
\mathcal{R}^{\mathrm{SIC-IAN}}_2= \left\{  (\lambda_{1},\lambda_{2}): \frac{\lambda_2}{\mathrm{Pr} \left\lbrace\mathrm{SNR}_{2} \geq \gamma_2 \right\rbrace}+ \right. \\
\left. + \frac{\mathrm{Pr} \left\lbrace\mathrm{SNR}_{2} \geq \gamma_2\right\rbrace-\mathrm{Pr} \left\lbrace\mathrm{SINR}_{2} \geq \gamma_2 \right\rbrace}{\mathrm{Pr}\left\lbrace\mathrm{SNR}_{2} \geq \gamma_2\right\rbrace\ \mathrm{Pr}\left\lbrace \left\lbrace \mathrm{SINR}_{21} \geq \gamma_2 \right\rbrace \cap \left\lbrace \mathrm{SNR}_1 \geq \gamma_1 \right\rbrace \right\rbrace}\lambda_1 < 1, \right. \\
 \left. \lambda_1 < \mathrm{Pr}\left\lbrace \left\lbrace \mathrm{SINR}_{21} \geq \gamma_2 \right\rbrace \cap \left\lbrace \mathrm{SNR}_1 \geq \gamma_1 \right\rbrace \right\rbrace \right\}
\end{aligned}
\end{equation}
}


Note that the stability region $\mathcal{R}^{\mathrm{SIC-IAN}}$ is concave/convex if $\mathrm{Pr}\left\lbrace \left\lbrace \mathrm{SINR}_{21} \geq \gamma_2 \right\rbrace \mid \left\lbrace \mathrm{SNR}_1 \geq \gamma_1 \right\rbrace \right\rbrace + \frac{\mathrm{Pr}\left\lbrace\mathrm{SINR}_2 \geq \gamma_2 \right\rbrace}{\mathrm{Pr}\left\lbrace\mathrm{SNR}_2 \geq \gamma_2 \right\rbrace} \gtrless 1$.

\section{Conclusions} \label{sec:conclusions}
We derived the stability region of the two-user interference channel for the general case and for different interference management strategies, namely treating interference as noise and successive interference cancelation at the receivers. Furthermore, we provided conditions for the convexity/concavity of the stability regions, as well as for which a certain interference management technique leads to broader stability region compared to the others.
Future work will include the closure of the presented regions for all powers/rates and the investigation of other techniques, such as joint decoding and interference alignment.

\bibliographystyle{IEEEtran}
\bibliography{thesis}

\end{document}